\documentclass[twocolumn,aps,prl,superscriptaddress]{revtex4-2}
\usepackage{mathtools}
\usepackage{bm}
\usepackage{amsmath}
\usepackage{amssymb}
\usepackage{graphicx}

\makeatletter
\usepackage{dcolumn}
\usepackage{bm}
\usepackage{multirow}

\usepackage{multirow}

\usepackage{bm}

\usepackage{color}
{\catcode `\@=11 \global\let\AddToReset=\@addtoreset}
\AddToReset{equation}{section}

\newcounter{mnotecount}[section]

\usepackage{chngcntr}
\counterwithout{equation}{section}

\makeatother

\begin{document}

\title{Probing modified gravity with magnetically levitated resonators}

\author{Chris Timberlake}
\email{C.J.Timberlake@soton.ac.uk}
\affiliation{Department of Physics and Astronomy, University of Southampton, Southampton
SO17 1BJ, UK}

\author{Andrea Vinante}
\email{anvinante@fbk.eu}
\affiliation{Istituto di Fotonica e Nanotecnologie – CNR and Fondazione Bruno Kessler, I-38123 Povo, Trento, Italy}

\author{Francesco Shankar}
\email{F.Shankar@soton.ac.uk}
\affiliation{Department of Physics and Astronomy, University of Southampton, Southampton
SO17 1BJ, UK}

\author{Andrea Lapi}
\email{lapi@sissa.it}
\affiliation{SISSA, Via Bonomea 265, 34136 Trieste, Italy}

\author{Hendrik Ulbricht}
\email{H.Ulbricht@soton.ac.uk}
\affiliation{Department of Physics and Astronomy, University of Southampton, Southampton
SO17 1BJ, UK}

\date{\today}
\begin{abstract}

We present an experimental procedure, based on Meissner effect levitation of neodymium ferromagnets, as a method of measuring the gravitational interactions between mg masses. The scheme consists of two superconducting lead traps, with a magnet levitating in each trap. The levitating magnets behave as harmonic oscillators, and by carefully driving the motion of one magnet on resonance with the other, we find that it should be easily possible to measure the gravitational field produced by a 4~mg sphere, with the gravitational attraction from masses as small as 30~$\mu$g predicted to be measurable within realistic a realistic measurement time frame. We apply this acceleration sensitivity to one concrete example and show the ability of testing models of modified Newtonian dynamics.

\end{abstract}
\maketitle

In 1778, Henry Cavendish published results which measured the gravitational interaction between masses in a laboratory setting, where the gravitational force between lead spheres caused a torsion balance to twist~\cite{Cavendish1798}. Since Cavendish's pioneering experiment, our understanding of gravity has improved dramatically. However, there are still a number of outstanding problems with regards to modern physics; gravity does not fit within the standard model~\cite{Gaillard1999} and there has been no conclusive evidence which links gravity with quantum mechanics~\cite{Penrose2014, Oriti2009, Kiefer2006}. In addition, modifications to Newton's second law of dynamics $F=ma$, especially in the regime of low accelerations such as in the MOdified Newtonian Dynamics (MOND) theory~\cite{MOND}, have been proposed as valid alternatives to dark matter~\cite{Zwicky} to explain, among other observables, the flat rotation curves of galaxies.  In the astrophysical community MOND is considered a viable possibility to reproduce a number of galactic observables~\cite{Brouwer2021}. Recently, Milgrom has claimed that MOND can reproduce the full scaling of the angular momentum of disc galaxies as a function of galaxy mass~\cite{Milgrom2021}. MOND is an example of a gravity theory that resembles Newtonian gravity only above a characteristic acceleration of $10^{-10}\ \text {m}/\text {s}^2$, but it significantly diverges at lower accelerations. Torsion pendulum experiments have confirmed Newton's second law down to $10^{-14}\ \text {m}/\text {s}^2$~\cite{Gundlach2007} using the restoring torque, and $10^{-12}\ \text {m}/\text {s}^2$~\cite{Little2014} for gravity based experiments. To fully test MOND, gravitational accelerations must be utilized. Klein has recently argued that these torsion pendulum experiments are compatible with MOND~\cite{Klein2020}. Given the open debate on the universal applicability of Newton’s gravitational law from the microscopic to galactic scales, i.e. low acceleration regimes, and given our still partial understanding of gravity in general, it is vital to continue testing MOND-like gravitational theories in a more manageable and controlled laboratory setting via experiments designed differently from a Cavendish-like one.

Since Einstein's general theory of relativity was devised in 1915~\cite{Einstein2019}, numerous experiments have confirmed these predictions on the astronomical scale, such as with gravitational lensing~\cite{Walsh1979}, gravitational redshift~\cite{Holberg2010} and more recently with the detection of gravitational waves~\cite{Abbott2016}. In laboratory based experiments, there has also been great progress in gravitational experiments, including tests of the equivalence principle~\cite{Ritter1990, Adelberger2001, Rosi2017}, non-Newtonian gravitational theories at short length scales~\cite{Geraci2015, Tan2020} and general relativistic effects~\cite{Asenbaum2017, Chou2010}. Additionally, proposals have been made which seek to test the potential quantum nature of gravity~\cite{Bose2017, Carlesso2019b}. There has also been a push to measure the gravitational attraction generated by small masses; to date, the smallest mass which has been experimentally demonstrated to cause a measurable gravitational force is approximately 90~mg~\cite{Westphal2021}.

Recently, levitated oscillators have shown great promise as sensors. By being free from mechanical clamping, levitated oscillators do not suffer from a large source of mechanical dissipation which limits conventional resonators. In particular, magnetically levitated oscillators have been proposed as very high quality factor oscillators, with extremely promising force and acceleration sensitivities~\cite{Prat-Camps2017}. Magnetic levitation has no active fields, which is often a dissipation source in optical~\cite{Li2010, Gieseler2012, Vovrosh2017} or electrical~\cite{Alda2016, Bullier2020} levitation. Levitating magnets in superconducting traps have been predicted to achieve acceleration sensitivities of $\sim 10^{-13}\ \text {m}/\text {s}^2/\sqrt{\text{Hz}}$~\cite{Prat-Camps2017}, where the magnet is a 10~mm diameter neodymium sphere. Experimentally, acceleration sensitivities of $1.2 \pm 0.2 \times 10^{-9} \ \text {m}/\text {s}^2/\sqrt{\text{Hz}}$ have shown, assuming a thermal noise limited system at 5~K, with 1~mm diameter neodymium spheres, with improvements in sensitivity anticipated by reducing the temperature of the thermal bath and installing vibration isolation~\cite{Timberlake2019}.

In this letter we propose an experiment which could measure the gravitational attraction of between two magnets levitated using the Meissner effect~\cite{Meissner1933}. The experimental procedure builds on the work shown in~\cite{Timberlake2019} and~\cite{Vinante2020}. The levitated magnets act as high quality factor mechanical resonators, and the gravitational attraction of one magnet, the source mass, will perturb the natural motion of other magnet, the test mass. The ultimate sensitivity of the levitating magnet is defined by the Brownian thermal force noise on the oscillator

\begin{equation}
    S_{FF}^{1/2} = \sqrt{\frac{4 k_B T m \omega_0}{Q}},
    \label{eq:Force_sensitivity}
\end{equation}

where $T$ is the temperature of the thermal bath the test mass is coupled to, $m$ is the mass of the test mass, $\omega_0$ is the oscillation frequency of the magnet and $Q$ is the quality factor, which is defined as $Q = \omega_0/\Gamma$ where $\Gamma$ is the damping rate of the oscillator. The thermal force noise can be considered the sensitivity, providing other noise sources are sufficiently suppressed. With respect to gravity, it is natural to discuss the acceleration sensitivity, rather than the force sensitivity. The acceleration sensitivity of a thermal noise limited mechanical oscillator is given by

\begin{equation}
    S_{aa}^{1/2} = \sqrt{\frac{4 k_B T \omega_0}{Qm}}.
    \label{eq:Acc_sensitivity}
\end{equation}

Over a certain measurement time $\tau$ this translates to a minimum measurable acceleration of ${a_{\text{min}} = \sqrt{4 k_B T \omega_0/Qm\tau}}$.

\begin{figure}[h!]
    \centering
    \includegraphics[width=0.9\linewidth]{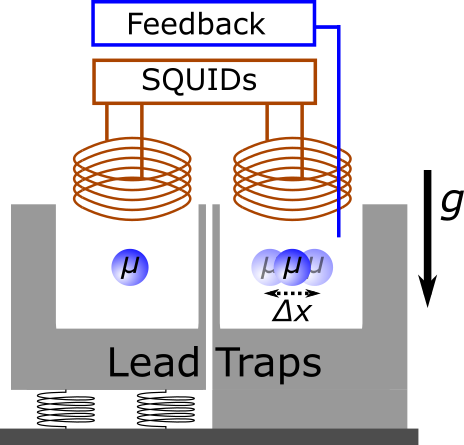}
    \caption{A schematic of the experimental setup. Neodymium magnet spheres are levitated in two independent lead wells, which are separated by a thin divide. Each trap also has a lead lid. The entire setup is cooled to around 300~mK in a helium-3 sorption refrigerator. Both levitating magnets have their position recorded using SQUIDS, and one of the magnets (source mass) has its motion driven, using electromagnetic forces, to an amplitude of $\Delta x/2$. This motion modulates the gravitational field experienced by the other magnet (test mass), which influences the motion and creates a detectable signal. The test mass trap is suspended by passive vibration isolation systems, which decouple the motion of the two traps from each other as well as isolate against background seismic noise. Earth's gravity is shown by $g$.}
    \label{fig:schematic}
\end{figure}

The experimental procedure we propose consists of two superconducting magnetic traps, similar to those seen in~\cite{Vinante2020}, separated by a thin superconducting wall. Each trap is fully enclosed with lead, which completely shields it electromagnetically from the other. We will then drive the motion of the source mass magnet at a specific frequency, on or near the natural resonance of the test mass magnet, in order to modulate the strength of the gravitational acceleration experienced by the test mass magnet. The mode of choice is perpendicular to the Earth's gravitational field. A schematic of this setup can be seen in Fig.~\ref{fig:schematic}. The two magnets are 1~mm diameter neodymium spheres~\cite{Timberlake2019} of mass 4~mg, separated by a center-of-mass distance of 7.5~mm, levitated in two identical superconducting traps. 4~mg magnets are chosen as a trade off between absolute acceleration sensitivity of the test mass, which increases as the mass increases, and the amplitude of motion and motional frequency, which both decrease as mass increases. If the amplitude is too small, it becomes too challenging to measure the motion. The traps are made of the type-I superconductor lead, and are mounted inside a 300~mK helium-3 refrigerator. The source mass can be driven using AC electric or magnetic fields, which will excite the test mass and cause an on-resonance signal enhancement\cite{Timberlake2019}. The motion of both magnets is measured using SQUIDs with a pick-up loop close to the equilibrium position of the levitated magnets.

\begin{figure}[h!]
    \centering
    \includegraphics[width=\linewidth]{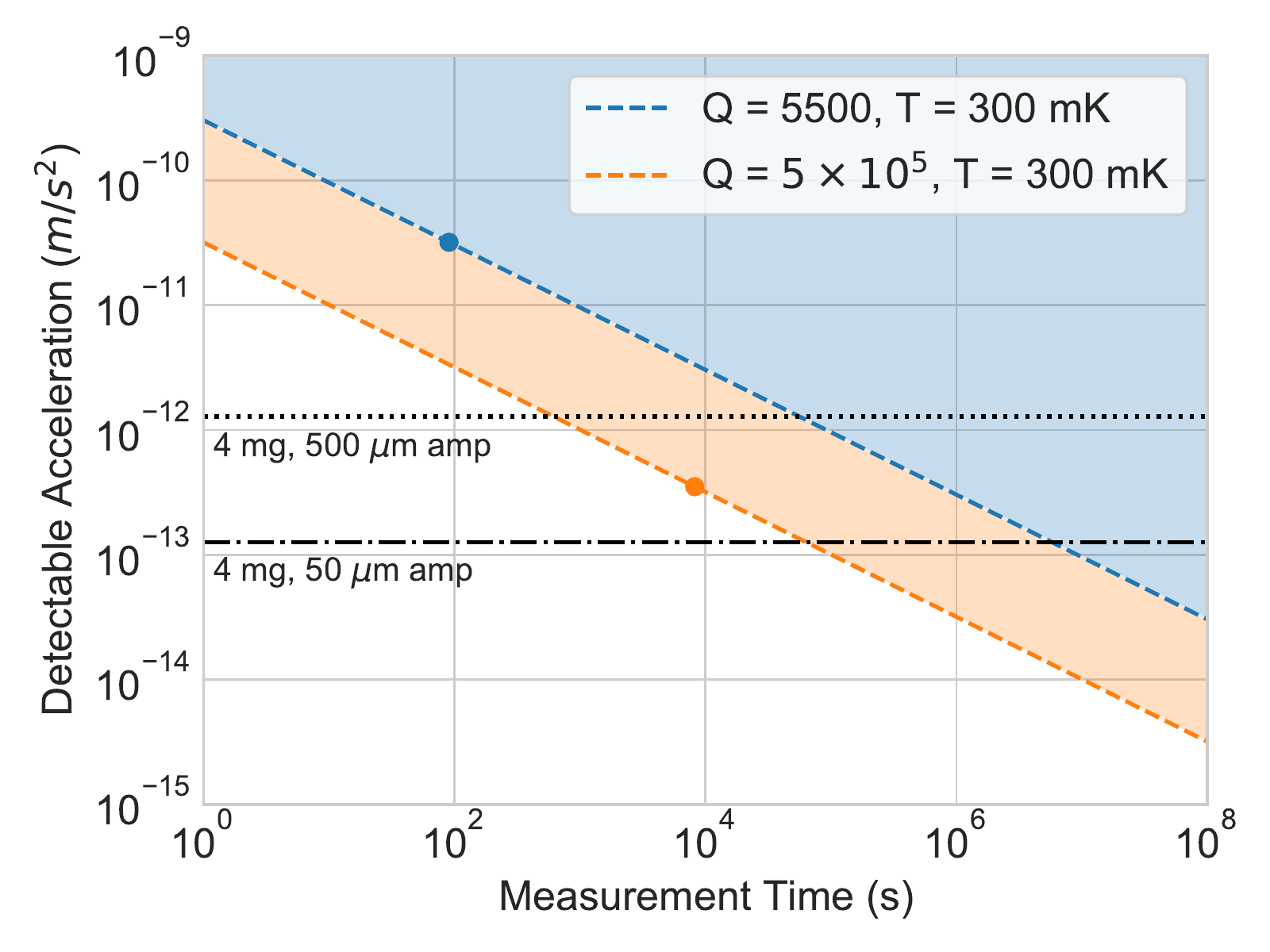}
    \caption{Figure showing the expected detectable acceleration as a function of measurement time for two realistic experimental scenarios. The blue dashed line represents the minimum detectable acceleration for Q-factor of $Q$ = 5500 (experimentally measured with 0.5~mm radius neodymium magnet~\cite{Timberlake2019}). The blue region region represents accelerations that can be measured with these experimental parameter. The orange dashed line shows the minimum detectable acceleration for $Q = 5 \times 10^5$, which is the predicted Q-factor with magnetic shielding. The orange region represents accelerations which can be measured with these experimental parameters. The blue and orange dots show the minimum measurement time required to resolve the motion of the oscillator for each Q-factor. The acceleration modulation that will be generated by driving the source mass magnet at an amplitude of $\Delta x/2 = 500~\mu$m and 50~$\mu$m is shown with the dotted and dashed-dotted horizontal lines.}
    \label{fig:acceleration}
\end{figure}

In a previous study~\cite{Timberlake2019}, the acceleration sensitivity of such a neodymium magnet sphere at a temperature of 5~K was found to be $S_a^{1/2} = 1.2 \pm 0.2 \times 10^{-9} \ \text {m}/\text {s}^2/\sqrt{\text{Hz}}$, for a thermal noise limited system. This corresponded to a Q-factor of $Q$ = 5500, which was limited by the lack of magnetic shielding around the magnet in the geometry of this experiment. In another complementary study, it was found that $Q > 10^7$ could be achieved by levitating micromagnetic neodymium particles within a lead trap~\cite{Vinante2020}, for librational degrees of freedom, and $Q > 10^6$ for translational modes. It was also apparent that the Q-factor that could be achieved was dependent on the radius on the magnet which is being levitated. The mechanism for this Q-factor limit is magnetic hysteresis losses within the magnet itself, and the relationship between Q-factor and magnet size is given by

\begin{equation}
    \frac{1}{Q} \propto \left(\frac{r}{z_0}\right)^3,
\end{equation}

where $r$ is the radius of the magnet and $z_0$ is the levitation height above the superconductor~\cite{Vinante2020}. By using this scale factor, we can estimate that the Q-factor achievable for a 0.5~mm radius neodymium magnet, which is fully shielded from external fields, will be $Q \sim 5 \times 10^5$. In Fig.~\ref{fig:acceleration}, the minimum detectable acceleration for a thermal noise limited system as a function of measurement time is plotted for the levitated magnet. Modulation amplitudes of $\Delta x/2 = 500~\mu$m and $\Delta x/2 = 50~\mu$m are considered. 

The calculations shown so far are based on the assumption that the oscillator is thermal noise limited in its motion. In order to ensure that the oscillator motion is indeed dominated by thermal noise, it's important to consider other potential sources of noise, such as thermal noise due to gas collisions, vibrational noise and other sources of magnetic noise. For calculations, we will assume an oscillation frequency of $\frac{\omega_0}{2\pi}$~=~30~Hz, which is consistent with what has been found with levitating 0.5~mm radius neodymium magnets~\cite{Timberlake2019}.

Firstly, we consider the noise due to collisions with gas particles within the experiment. In such a cryogenic experiment, the background gas is helium. In the helium-3 refrigerator, helium pressures of $10^{-7}$~mbar are easily obtainable at 300~mK, with pressures of $<10^{-10}$~mbar measured after baking out the entire cryostat insert. The damping due to gas collisions can be given by

\begin{equation}
    \Gamma_{\text{gas}} \approx \frac{15.8 r^2 P}{m v_{\text{gas}}},
    \label{eq:gas_damping}
\end{equation}

where $r$ is the radius of the levitated sphere, $P$ is the gas pressure and $v_{\text{gas}} = \sqrt{3 k_B T/m_{\text{gas}}}$ is the thermal velocity of the gas molecules with mass $m_{\text{gas}}$~\cite{Beresnev1990}. This corresponds to a displacement noise of $\sim3.8 \times 10^{-17}~\text{m}/\sqrt{\text{Hz}}$ at a frequency of $\frac{\omega_0}{2\pi}$~=~30~Hz. For this oscillation frequency, and our expected $Q = 5 \times 10^5$, the vibrational displacement noise will be $\sim1.5 \times 10^{-15}~\text{m}/\sqrt{\text{Hz}}$ for the magnetic hysteresis thermal noise, which is significantly higher than the thermal noise due to gas collisions calculated using a conservative estimate of gas pressure in the experimental chamber, meaning the damping effect of gas collisions is negligible. 

At low frequencies, the oscillator is susceptible to seismic noise vibrations. To reach the thermal noise limit, it is essential that seismic noise is below this level. In a typical laboratory setting, one might expect a seismic noise contribution of $\lvert10^{-9}/\left(f/1 \ \text{Hz}\right)^2\rvert~\text {m}/\sqrt{\text{Hz}}$ above 1~Hz~\cite{ze2001}, which for $\frac{\omega_0}{2\pi}$~=~30~Hz is $1.1 \times 10^{-12}~\text {m}/\sqrt{\text{Hz}}$.

Therefore, the vibrational noise must be reduced by approximately three order of magnitude around 30~Hz to reach the thermal noise limit. However, it is important to consider any cross-coupling between the two traps. By driving the motion of the source magnet inside the trap, we unwittingly will provide a force which shakes the entire trap on resonance, meaning that passive vibration isolation of the test mass trap, similar to those in Ref.~\cite{Leng2021, DeWit2019} is needed, but the source mass trap must be rigidly clamped to the Earth. Approximately 60~dB of isolation is needed for the test mass lead trap to reduce seismic noise below thermal noise. 

The next source of noise to consider is magnetic noise. Currently, we have assumed that the dominating noise source will be due to unavoidable magnetic hysteresis, but to ensure this is the case we must first mitigate the effects of other magnetic noise sources. Eddy current dissipation can't be completely mitigated, as the magnet itself is made of metal. However, the internal eddy current dissipation has been shown to significantly less than the magnetic hysteresis theoretically~\cite{Prat-Camps2017}, and also applied to real experimental finding where neodymium magnets are levitated~\cite{Vinante2019}. One caveat is the metallic coating on commercial magnets will add a significant amount of eddy current damping~\cite{Timberlake2019}, so it may be necessary to levitate uncoated neodymium magnets. Stray magnetic fields can significantly increase the damping experienced by a levitated magnet~\cite{Vinante2019}. A magnetic shield can be placed around the superconducting traps to mitigate this. Additionally, the thickness of the lead divide between the two traps is 0.5~mm, which is far larger than the London penetration depth (37~nm) and the coherence length (83~nm) of lead~\cite{Kittel1996}. This means that any electromagnetic interaction between the two traps will be screened by the divide, which would otherwise overwhelm the gravitational signal. 

Given the sensitivities shown in Fig.~\ref{fig:acceleration}, it may be possible to measure smaller accelerations with a similar experiment. In this experimental design, the source mass would have a smaller mass than the test mass. The 300~mK refrigerator can be held at 300~mK for a hold time of approximately 40~hours. Using this as the maximum measurement time, we find that the smallest measurable acceleration is $8 \times 10^{-14}~\text{m}/\text{s}^2$, for the 4~mg test mass with a quality factor of $Q = 5 \times 10^5$. This translates to a minimum source mass of 30~$\mu$g for 500~$\mu$m of driving.

\begin{figure}[h!]
    \centering
    \includegraphics[width=\linewidth]{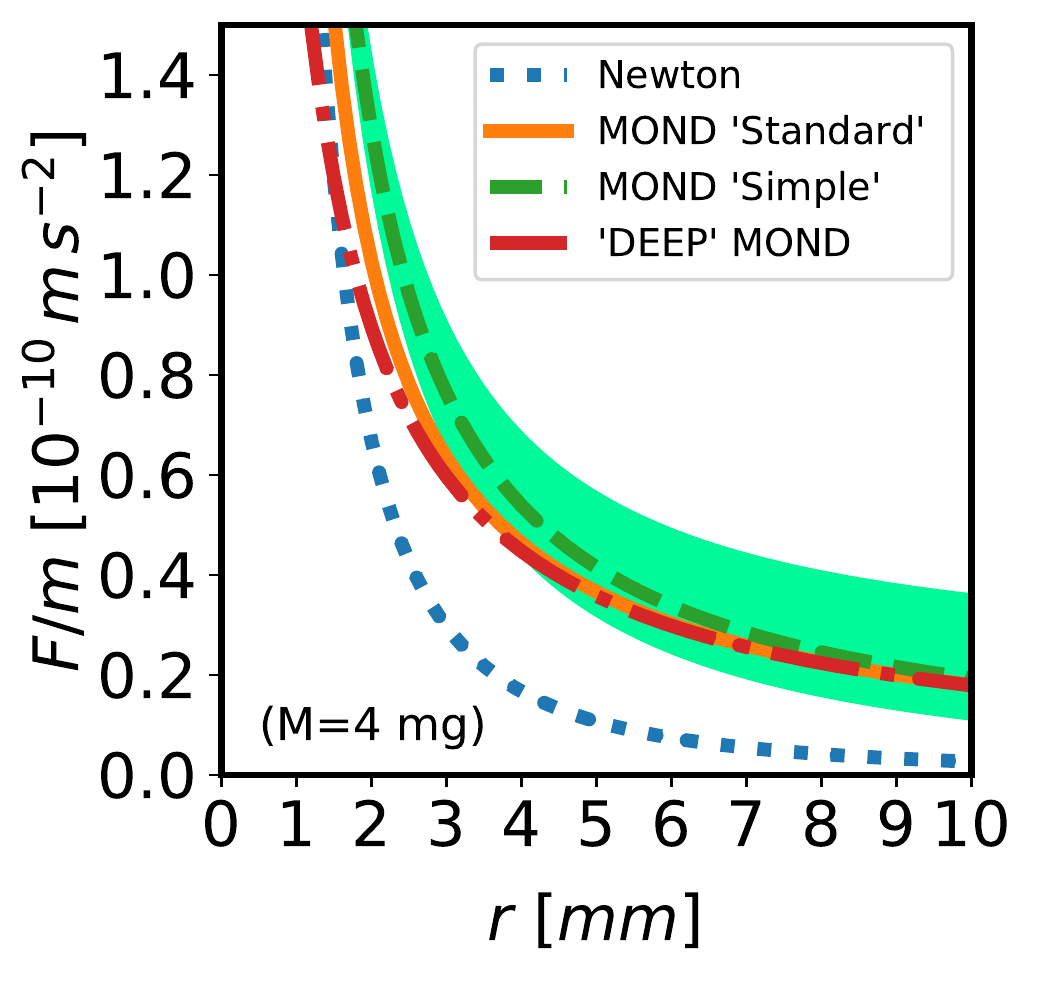}
    \caption{Figure presenting the predicted acceleration (in units of $10^{-10}\, \text {m}/\text {s}^2$) imprinted on a test particle by a particle of mass $M=4\, \text{mg}$, as a function of particle separation $r$ in mm. The blue, dotted line is the expected acceleration in gravitational Newtonian dynamics, whilst the green dashed and orange solid lines are the predicted gravitational accelerations expected in MOND based on Eq.~\ref{eq:MOND} (the green shaded region brackets the simple MOND model when including the EFE via Eq.~\ref{eq:EFE} with $e=\pm 0.1$, see text for details). The red dot-dashed line is the acceleration from Eq.~\ref{eq:MOND} in the DEEP MOND limit, i.e. when $\mu[x]=x$.}
    \label{fig:MOND}
\end{figure}

The estimated precision of our gravitational acceleration measurements will directly test Newton's second law. Here, we consider the constraints on departures from Newtonian dynamics for MOND-like phenomenologies as one example of fundamental physics that could be probed with our experiment. Fig.~\ref{fig:MOND} shows the expected acceleration (in units of $10^{-10}\, \text {m}/\text {s}^2$) on a test particle $m$ imparted by a particle of a mass $M=4~\text{mg}$ in Newtonian dynamics (blue, dotted line) and in MOND. The radial acceleration $a_r$ in MOND is linked to the Newtonian gravitational acceleration via the relation~\cite{MOND}

\begin{equation}
    m \mu \left[\frac{a_r}{a_0}\right]a_r=F=\frac{GMm}{r^2}\, ,
    \label{eq:MOND}
\end{equation}

where the function $\mu[x] \rightarrow x$ in the ``DEEP'' MOND regime at very low accelerations $x \ll 1$ (red dot-dashed line), whilst approaching unity, and thus reconciling $a_r$ to Newtonian dynamics, when $x \gg 1$, i.e. when the accelerations become larger than the threshold value $a_0=1.2\times 10^{-10}\, \text {m}/\text {s}^2$. The full functional form for $\mu[x]$ is usually given~\cite{Famaey2005} in a ``simple'' $\mu[x]=x/(1+x)$ or ``standard'' $\mu[x]=x/\sqrt{1+x^2}$ form (respectively, green dashed and orange solid lines). Fig.~\ref{fig:MOND} depicts a ``particle rotation curve'' which in MOND rapidly flattens out with increasing separation $r$, mimicking a weaker dependence of the acceleration on distance $a_r\propto 1/r$, in a similar fashion to what is observed in galactic rotation velocity curves at large distances from the galactic centres. By achieving a precision of $0.1 a_0\sim 10^{-11}\, \text {m}/\text {s}^2$ down to even $10^{-3} a_0\sim 10^{-13}\, \text {m}/\text {s}^2$, according to our estimates in Fig.~\ref{fig:acceleration}, we will be able to distinguish MOND from Newtonian dynamics at a high confidence level for any separation $r\gtrsim 3\, \text{mm}$ and for a large range of masses $M$. In addition, by repeating the experiment for different masses $M$, we can build the correlation between the test's particle velocity and central mass $M$, which MOND would predict to be extremely tight with a slope equal to 4~\cite{MOND}. Exploring different realizations of the same experiment and providing predictions in terms of both accelerations and velocities allows to better control systematics in either variable~\cite{Gundlach2007, Das2013}. 

We also note that in MOND the internal dynamics of a system is not independent of external fields (the so-called ``External Field Effect'', EFE). From Eq. 59 of ~\cite{Famaey12} we have found that, for the simple MOND case, the addition of an external field $g_{\rm ext}$ just amounts to replace in Eq.~\ref{eq:MOND} the interpolating function $\mu[x]$ with 
\begin{equation}
\mu_e[x] = \frac{x}{1+x+e}\,\left[1+\frac{e}{x}\,\frac{2+e}{1+e}\right]
\label{eq:EFE}
\end{equation}

where $e\equiv a_{\rm ext}/a_0$. By design, our experiment can be locally considered as an inertial frame, as the magnetic levitation perfectly cancels Earth's gravitational field. However, a residual, cumulative EFE could still be detected from external galaxies. From the modelling of 153 SPARC rotating galaxies, \cite{Chae20} estimated an average value of $e\sim 0.033$ when including the EFE. We show in Fig.~\ref{fig:MOND} the effect of including the EFE via Eq.~\ref{eq:EFE} and a value of $e=\pm0.1$ (green shaded region). Despite adopting a value of $e$ about three times higher than the one estimated by \cite{Chae20}, we still find that the a deviation from MOND could still be detected at high significance.  

Our results, in turn, will have profound consequences on many aspects of fundamental physics, and specifically on our understanding of gravity, inertial motions, general relativity, particle physics and cosmology, further supporting -- or challenging -- the existence of the elusive dark matter component in and around galaxies. 

In summary, we have introduced an experiment based on superconducting levitation of ferromagnets to study gravitation interactions between two levitating masses. This technique opens up the possibility of measuring gravitational interaction between mg masses, with predictions based on pragmatic calculations based on previous experimental works~\cite{Timberlake2019, Vinante2020}. With the sensitivities explored here, it may be possible to measure gravitational accelerations produced by sub-mg masses in future iterations of this experimental design. Such experiments could also be used to provide an alternative measure of the gravitational constant ``big G". Our proposed experiment offers an alternative to Cavendish-like torsion pendulum gravitational experiments, and has the potential to measure departures from Newtonian dynamics at high confidence levels, setting unique constraints on MOND-like phenomenological models (see Fig.~\ref{fig:MOND}), vital for cosmology and particle physics, as well as, when adapted in scale and mass, on Yukawa-like potentials~\cite{Tan2020} to probe predictions beyond the standard model. 

The authors would like to acknowledge the Leverhulme Trust {[}RPG-2016-046{]} and the EU Horizon 2020 research and innovation programme under grant agreement No 766900 {[}TEQ{]} for funding support.

\bibliographystyle{apsrev4-1}

%

\appendix
\end{document}